\documentclass[runningheads,envcountsame]{llncs}

\usepackage{afterpage}
\usepackage{amsfonts}
\usepackage{amsmath}
\usepackage{amssymb}
\usepackage{enumerate}
\usepackage[utf8]{inputenc}
\usepackage{chemformula}
\usepackage[version=4]{mhchem}
\usepackage{longtable}
\usepackage{listings}
\usepackage{url}
\lstdefinelanguage{Maple}%
{morekeywords={and,assuming,break,by,catch,description,do,done,%
elif,else,end,error,export,fi,finally,for,from,global,if,%
implies,in,intersect,local,minus,mod,module,next,not,od,%
option,options,or,proc,quit,read,return,save,stop,subset,then,%
to,try,union,use,uses,while,xor},%
sensitive=true,%
morecomment=[l]\#,%
morestring=[b]",%
morestring=[d]"%
}[keywords,comments,strings]
\usepackage[]{todonotes}
\usepackage{graphicx}

\usepackage{algorithm,algorithmic}

\newcommand{\CC}{\mathbb{C}}

\newcommand{\KK}{\mathbb{K}}
\newcommand{\NN}{\mathbb{N}}
\newcommand{\QQ}{\mathbb{Q}}
\newcommand{\RR}{\mathbb{R}}

\newcommand{\LL}{\mathbb{L}}

\newcommand{\RGB}{\operatorname{ReducedGroebnerBasis}}
\newcommand{\PGBMAIN}{\operatorname{PGBMAIN}}
\newcommand{\MDBasis}{\operatorname{MDBasis}}

\spnewtheorem{conf}[theorem]{Conjecture}{\bfseries}{\upshape}
\spnewtheorem{rem}[theorem]{Remark}{\bfseries}{\upshape}
\newtheorem{Example}{Example}

\begin{document}
\title{Testing Binomiality of Chemical Reaction Networks Using
  Comprehensive Gr\"obner Systems}

\titlerunning{Parametric Binomiality of Steady State Ideals}

\author{Hamid Rahkooy\inst{1}
  \and Thomas Sturm\inst{2,1,3}
}

\institute{
  MPI Informatics, Saarland Informatics Campus, Saarbrücken, Germany\\
  \and
  CNRS, Inria, and the University of Loraine, Nancy, France\\
  \and
  Saarland University, Saarland Informatics Campus, Saarbrücken,
  Germany\\[1ex] 
  \email{hamid.rahkooy@mpi-inf.mpg.de},
  \email{thomas.sturm@loria.fr}%
}

\maketitle

\begin{abstract}
  We consider the problem of binomiality of the steady state ideals of
  biochemical reaction networks. We are interested in finding
  polynomial conditions on the parameters such that the steady state
  ideal of a chemical reaction network is binomial under every
  specialisation of the parameters if the conditions on
  the parameters hold.  We approach the binomiality problem using
  Comprehensive Gr\"obner systems. Considering rate constants as
  parameters, we compute comprehensive Gr\"obner systems for various
  reactions. In particular, we make automatic computations on n-site
  phosphorylations and biomodels from the Biomodels repository using
  the grobcov library of the computer algebra system Singular.
  
\end{abstract}

\section{Introduction}\label{sec:intro}
  A \textit{chemical reaction} is a transformation between two sets of
  chemical objects called chemical \textit{complexes}. The objects
  that form a chemical complex are chemical \textit{species}. In fact,
  complexes are formal sums of chemical species representing the left
  and the right hand sides of chemical reactions. A \textit{chemical
    reaction network} is a set of chemical reactions. For example
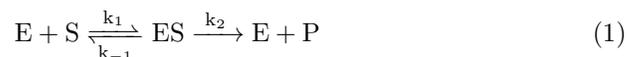
\begin{equation}\label{crn1}
  \ce{E + S <=>[k_1][k_{-1}] ES ->[k_2] E + P}
\end{equation}
is a chemical reaction network with one reversible reactions and one
non-reversible reaction. This reaction network is a well-known
network, called the {\em Michaelis--Menton} reaction network.

A \textit{kinetics} of a chemical reaction network is an assignment of
a rate function to each reaction in the network. The rate function
depends on the concentrations of the species. A kinetics for a
chemical reaction network is called \textit{mass-action} if for each
reaction in the network, the rate function is a monomial in terms of
the concentrations of the species, such that the exponents are given
by the numbers of molecules of the species consumed in the reaction,
multiplied by a constant called {\em rate constant}. In the
Michaelis--Menton reaction, $k_{1}$, $k_{-1}$, $k_{2}$ are the rate
constants. In this article, we assume mass-action kinetics.

A system of autonomous ordinary differential equations can be used to
describe the change in the concentration of each species over time in
a reaction. For example, in the Michaelis--Menton reaction, let the
variables $s, p, c, e$ represent the concentrations of the species
$S, P, ES, E$ respectively.  The ordinary differential equations
(ODEs) describing change of the concentrations of the species for this
reaction network are the following:
\begin{align}
  \dot{s} &= f_s =  -k_{1} se + k_{-1} c, \label{ode1}\\
  \dot{p} &= f_p =  k_{2} c, \label{ode2} \\
  \dot{c} &= f_c =  k_{1} s e - (k_{-1}+k_{2}) c, \label{ode3} \\
  \dot{e} &= -f_c . 
\end{align}

Solutions of the polynomials $f_s$, $f_p$, $f_c$ and $-f_c$ give us
the concentrations of the species in which the system is in
equilibrium. In fact, the solutions of $f_s$, $f_p$, $f_c$ and $-f_c$
are called the steady states of the chemical reaction network.
Accordingly, the ideal generated by $f_s$, $f_p$, $f_c$ and $-f_c$,
i.e.,
$I=\langle f_s, f_p, f_c,-f_c \rangle \subseteq
\KK[k_1,k_{-1},k_2][s,p,c,e] $, where $\KK$ is a field, is called the
{\em steady state ideal} of the Michaelis--Menton network.
For a thorough introduction on chemical reaction network theory, 
refer to Feinberg's Book \cite{feinberg-book} and his lecture notes
\cite{Feinberg-Lectures}. We follow the notation of Feinberg's book in
this article.

A {\em binomial ideal} is an ideal that is generated by a set of
binomials.  In this article, we consider the problem of binomiality of
steady state ideals when the rate constants are specialised over a
field extension of $\KK$, that is, when the rate constants have been
assigned values from an extension of $\KK$, typically the closure of
$\KK$. More precisely, we are interested in conditions over the rate
constants (typically given by polynomial equations on rate constants),
such that for every values of the rate constants in the extension
field, the steady state ideal is binomial under those conditions. In
this article, we often use parameters instead of rate constants, an
indication that they can be specialised. Therefore, we consider the
{\em parametric binomiality problem}.

Let us consider the  steady state ideal of the Michaelis--Menton reaction:
\begin{equation}
  I = I=\langle f_s, f_p, f_c
  \rangle \subseteq
  \KK[k_1,k_{-1},k_2][s,p,c,e],
\end{equation}
given by Equations (\ref{ode1})--(\ref{ode3}).  One can observe that
$f_c=-f_s+f_p$. Hence, $I = \langle f_s,f_p\rangle$. Having fixed the
term ordering induced by $c>s>e$, one may consider further reducing
$f_s$ by $f_p$, i.e., $f_s-f_p = (k_{-1}-k_1)c -k_1 se$. As the rate
constants in a chemical reaction take values, $k_{-1}-k_1$ may
vanish. In this case, if the leading term of $f_s-f_p$ vanishes, then
it will be a monomial, and therefore, the reduced Gr\"obner basis of
$I$ will be the monomial ideal generated by $\{ k_2c,-k_1 se\}$, given
that $k_2 \ne 0$ and $k_{-1} \ne 0$. This example shows that the
Gr\"obner basis of the steady state ideal (and the steady states of
the reaction) can change depending on the values of the rate
constants. Therefore, we must consider distinct cases for the
parameters when analysing a reaction network. Thinking purely in terms
of computer algebra, this example illustrates the idea behind {\em
  Comprehensive Gr\"obner bases}.  In this article, we investigate the
conditions on the parameters of a steady state ideal (or equivalently
on the rate constants of a reaction) such that the steady state ideal
is binomial when those conditions on the parameters hold.

In the literature, a slightly different notions of binomiality has
been considered. Eisenbud and Sturmfels in
\cite{eisenbud-sturmfels-binomials} call an ideal binomial if it is
generated by polynomials with at most two terms. Some authors, e.g.,
P\'erez-Mil\'an et al. \cite{perez_millan_chemical_2012}, have studied
the binomiality of steady state ideals according to the definition in
\cite{eisenbud-sturmfels-binomials}. However, in this article, our
definition does not include those ideals that include monomials. This
difference in the definition, obviously, affects the steady state
variety of binomial chemical reaction networks in practice.

Binomial ideals and toric varieties have rich history in chemical
reaction networks theory. Binomiality corresponds to detailed balance,
which is a very important concept in thermodynamics. Detailed balance
means that at thermodynamic equilibrium, the forward and backward
rates should be equal for all reactions. Detailed balance has been
historically used by Einstein \cite{einstein1916strahlungs},
Wegscheider \cite{Wegscheider1901} and by Onsager
\cite{onsager1931reciprocal}.  Some of the subsystems of molecular
devices can satisfy binomiality conditions. Another interesting point
to study binomiality is because the analysis of properties such as
multi-stationarity and stability are easier to establish for binomial
systems. Toricity, also known as complex, or cyclic, or semi-detailed
balance is also known since Boltzmann that has used it as a sufficient
condition for deriving his famous H-theorem
\cite{boltzmann1964lectures}. Toricity implies binomiality, but the
converse is not true. A toric variety is indeed irreducible, however a
binomial steady state ideal may have an irreducible variety, which
would not be toric. However, every variety of a binomial ideal
includes a toric variety as its irreducible component. A toric system
must obey constraints on the rates constants, such as the well known
Weigscheider---Kolmogorov condition, which implies the equality of the
products of forward and backward rates constants in cycles of
reversible reactions.

Mathematicians have considered binomiality and toricity and
investigated their properties thoroughly, among them existing
literature are the work by Fulton \cite{fulton_introduction_2016},
Sturmfels \cite{sturmfels_grobner_1996} and Eisenbud et
al.~\cite{eisenbud-sturmfels-binomials}. Binomiality implies
\textit{detailed balancing} of reversible chemical reactions, which
has been studied by Gorban et
al.~\cite{gorban_generalized_2015,gorban_three_2015} and Grigoriev and
Weber \cite{GrigorievWeber2012a}. Toric dynamical systems have been
studied by Feinberg \cite{Feinberg1972} and Horn and Jackson
\cite{Horn1972}.  Over the real numbers Craciun et al.~have studied
the toricity problem in \cite{craciun_toric_2009}. In the latter work,
it hs been shown that \textit{complex balanced systems} are the same
as toric dynamical systems, although \textit{toric steady states} are
different from that.  Binomiality implies much simpler criteria for
multistationarity \cite{dickenstein2019,sadeghimanesh2019}.

P\'erez-Mil\'an, et al. presented a sufficient linear algebra
conditions with inequalities for binomiality of the steady state
ideals \cite{millan2012chemical}. The idea in the latter has been
developed in \cite{millan_structure_2018}, where MESSI reactions have
been introduced. Conradi and Kahle have proved in
\cite{conradi2015detecting} that for homogenous ideals the latter
sufficient condition is necessary as well, and introduced an algorithm
for that case. Their algorithm has been implemented in Maple and
Macaulay II in \cite{MapleCK,alex2019analysis}.  A geometric view
towards toricity of chemical reaction networks has been given by
Grigoriev et al.~in \cite{grigoriev2019efficiently}, where shifted
toricity has been introduced, algorithms presented for testing shifted
toricity and complexity bounds and experimental results are
discussed. In \cite{grigoriev2019efficiently}, the two main tools from
computer algebra,, quantifier elimination
\cite{DavenportHeintz:88a,Grigoriev:88a,Weispfenning:88a} and
Gr\"obner bases
\cite{Buchberger:65a,bb-system,Faugere:99a,Faugere:02a} are used. Also
recently, the authors introduced a first order logic test for toricity
\cite{sturm2020firstorder}. An efficient linear algebra method for
testing unconditional binomiality has been presented in
\cite{DBLP:conf/casc/RahkooyRS20} and a graph-theoretical equivalent
of the method is given in \cite{DBLP:conf/synasc/RahkooyM20}.

Testing binomiality of an ideal is a difficult problem, both from a
theoretical and a practical point of view. A typical method to test
binomiality is via computing a Gr\"obner basis. It has been shown that
computing a G\"obner basis is EXPSPACE-complete
\cite{mayr_complexity_1982}, which shows the difficulty of the
binomiality problem from the computational point of view. The approach
proposed for testing binomiality of steady state ideals in
\cite{millan2012chemical,conradi2015detecting} relies on linear
algebra. In this approach the computations are done without
considering the values of the parameters. Also large matrices are
constructed in this approach.

Existing work on binomiality of chemical reaction networks typically
ignores specialisation of the parameters, often treating them as
variables and carrying on the computations. For instance, fixing an
ordering in which the parameters are smaller than the variables, e.g.,
lexicographic ordering, one may consider computing a Gr\"obner basis
of the steady state ideal and then eliminating the variables. Then the
elimination ideal will be in the ring of parameters and may result in
conditions on the parameters such that the original ideal is
binomial. However, this approach does not consider the fact that in
the process of computations, some terms can be vanished, if parameters
are specialised.
  
In contrast, our approach is to use comprehensive Gr\"obner bases,
which considers specialisations of the parameters. A comprehensive
Gr\"obner basis of an ideal is a finite set of polynomials on the
parameters and the variables, such that it is a Gr\"obner basis under
every value assignment in the parameters. Therefore, a steady state
ideal is binomial if its comprehensive Gr\"obner basis is
binomial. This observation reduces testing binomiality of a steady
state ideal under specialisation into testing binomiality of a
comprehensive Gr\"obner basis. Computing a comprehensive Gr\"obner
basis results in a partitioning of the ambient space into certain
varieties and computations of certain set of polynomials associated to
each of those varieties, such that if the parameters are specialised
from the variety, the associated polynomial set is a Gr\"obner
basis. Such a partition with its associated polynomial sets is called
a Gr\"obner system.
%
  Computing comprehensive Gr\"obner bases is at least as difficult as
  computing Gr\"obner bases. Hence, testing binomiality via
  comprehensive Gr\"obner bases is a hard problem.

  The concept of comprehensive Gr\"obner bases has been introduced by
  Weispfenning in his seminal work \cite{Weispfenning92}. He later
  introduced canonical comprehensive Gr{\"{o}}bner bases in
  \cite{Weispfenning03}. A source of introduction to comprehensive
  Gr\"obner basis is Becker and Weispfenning's book
  \cite{Weispfenning-Becker93}. Weispfenning also worked on the
  relation between comprehensive Gr{\"{o}}bner bases and regular rings
  \cite{Weispfenning06}.  Later, several authors worked on the topic
  and introduced more efficient algorithms and polished the theory of
  comprehensive Gr\"obner bases. Suzuki-Sati's approach to Gr\"obner
  bases is presented in \cite{SuzukiS03}. Montes has worked
  extensively on comprehensive Gr\"obner bases, introduced several
  algorithms and developed the theory \cite{Montes02,DarmianHM11}. In
  particular, Montes' book, the Gr\"obner Cover \cite{grobcov-book} is
  a great source for computations, among other interesting aspects,
  that can be used as a guide to the Singular library grobcov.lib
  \cite{singular} for computing comprehensive Gr\"obner bases. Among
  the most efficient algorithms for computing comprehensive Gr\"obner
  bases are the algorithms given by Kapur et
  al.\cite{KapurSW13a,Kapur17}. Dehghani and Hashemi studied Gr\"obner
  walk and FGLM for comprehensive Gr\"obner bases
  \cite{DarmianH17,HashemiDB17} and implemented several algorithms for
  computing comprehensive Gr\"obner bases and related topics in Maple
  \cite{HashemiDB17}.\footnote{\url{https://amirhashemi.iut.ac.ir/sites/amirhashemi.iut.ac.ir/files//file_basepage/pggw_0.txt}}

  To the best of our knowledge, to this date, comprehensive Gr\"obner
  bases have not been used in chemical reaction networks
  theory. Previous studies on binomiality of steady state ideals have
  considered Gr\"obner bases, linear algebra on stoichiometric
  matrices, etc., however, never have considered the change in the
  polynomials during computations when the values are assigned to the
  parameters. For instance, it is known that detailed balancing holds under some 
  polynomial conditions on the parameters.
  However, the fact that specialisation of the rate constants may
  affect the computations has not beed considered.  The authors'
  previous work on toricity
  \cite{grigoriev2019efficiently,sturm2020firstorder} considers the
  toricity problem when the parameters have already been assigned real
  positive values. Other articles of the authors have considered
  unconditional binomiality, that is, when the rate constants are
  considered variables
  \cite{DBLP:conf/casc/RahkooyRS20,DBLP:conf/synasc/RahkooyM20}. The
  present article is the original work that consideres specialisation
  of the parameters and uses comprehensive Gr\"obner bases to study
  the binomiality under specialisations.

  The plan of the article is as follows. Section \ref{sec:intro} gives
  an introduction to the necessary concepts of chemical reaction
  network theory, reviews the literature and presents the idea of the
  present article. Section \ref{sec:prelim} explains the preliminaries
  required on comprehensive Gr\"obner systems, explains the main
  concepts and sketches the idea behind computing comprehensive
  Gr\"obner bases.  Section \ref{sec:crn-cgs} includes the main
  computations, where we show our computations on $n-$phosphorylations
  and biochemical reactions and present the benchmarks. We furthermore
  compare our computations using comprehensive Gr\"obner bases with
  some earlier work on the binomiality problem that does not take into
  account the specialisation of the rate constants.  In Section
  \ref{sec:conclusion} we summarise our results and draw some
  conclusions.


\section{Preliminaries on Comprehensive Gr\"obner
  System}\label{sec:prelim}

We review the required definitions, theorems and an algorithm on
comprehensive Gr\"obner systems, mainly from the original work of
Weispfenning \cite{Weispfenning92} and Kapur, et al.'s work
\cite{KapurSW13a}.


Let $\KK$ be a field, $R =\KK[U] = \KK[u_1,\dots,u_m]$ be the ring of
polynomials over $\KK$ in the indeterminates $u_1,\dots,u_m$ and let
$S=\KK[U][X]=\KK[u_1,\dots,u_m][x_1,\dots,x_n]$ be the ring of
polynomials over $\KK[U]$ with the indeterminantes
$x_1,\dots,x_n$. Assume that $X \cap U = \emptyset$. We call
$u_1,\dots,u_m$ the parameters of the ring $S$ and $x_1,\dots,x_n$ the
variables of $S$. In fact, the coefficients of every polynomial in $S$
are themselves polynomials in parameters.  For every
$\alpha=(\alpha_1,\dots,\alpha_n) \in \NN^n$, by $X^\alpha$ we denote
$x_1^{\alpha_1}\dots x_n^{\alpha_n}$ and by $U^\alpha$ we denote
$u_1^{\alpha_1}\dots u_n^{\alpha_n}$.  In this paper, $\KK$ is either
$\RR$ or $\CC$.  By the variety of an ideal $I$ (or a set of
polynomials $F$), we mean the set of solutions of the ideal $I$ (or
the set of polynomials $F$) and we denote it by $V(I)$ (or $V(F)$).

Let $<_1$ and $<_2$ be term orders on $\KK[U]$ and $\KK[X]$,
respectively. We define a block order $<$ produced by the latter on
$\KK[U][X]$.  Firstly, define $u_i < x_j$ for all
$1\leq i \leq m, 1 \leq j \leq n$. Secondly, define
$X^{\alpha_1}U^{\beta_1} < X^{\alpha_2}U^{\beta_2}$ if either
$X^{\alpha_1} < X^{\alpha_2}$ or
$(X^{\alpha_1} = X^{\alpha_2} \land U^{\alpha_1} < U^{\alpha_2})$.  A
polynomial of the form $c_\alpha p(U)X^\alpha$, where
$\alpha \in \NN^n$, $c_\alpha \in \KK$ and $p(U) \in R$, is called a
term in $\KK[U][X]$. A monomial is a term of the form
$X^\alpha$. Leading monomial, leading term and leading coefficient of
the polynomials in $\KK[U][X]$ are defined with respect to the block
ordering $<$.
  
A specialisation of $S$ is a ring-homomorphism from the ring of
parameters $R=\KK[U]$ into some field $\LL$, i.e.,
$\sigma :R \rightarrow \LL$. Obviously $\KK$ is embedded in $\LL$. We
consider $\LL$ to be an algebraically closed field in this paper.
Every specialisation is uniquely determined by its restriction to
$\KK$ and its images on the parameters $u_1,\dots,u_m$ and vice
versa. A specialisation $\sigma : R \rightarrow \LL$ has a canonical
extension to a ring-homomorphism
$\bar\sigma :S \rightarrow \LL[x_1,\dots,x_n]$, i.e., for every
$f=\sum_{i \in I}a_i(U)X^{\alpha_i}, \bar\sigma(f)=\sum_{i \in
  I}\sigma(a_i(U)) X^{\alpha_i}$, where $a_i(U)\in R$ and
$X^{\alpha_i}$ is a monomial in $\KK[X]$. Following Weispfenning's
notation, we denote $\bar\sigma$ by $\sigma$ as well. Specialisation
of a set of polynomials $F$ by $\sigma$, denoted by $\sigma(F)$, is
defined to be the set of specialisations of the polynomials in
$F$. Accordingly, a specialisation of an ideal $I$ by $\sigma$ is
defined, and is denoted by $\sigma(I)$.  Following Kapur, et
al. \cite{KapurSW13a}, in this paper we only consider specialisations
induced by the elements $a \in \LL^m$, that is,
$\sigma_a:f \rightarrow f(a)$, where $f \in R$.

  Below we mention the definition of comprehensive Gr\"obner system
  and comprehensive Gr\"obner basis, which are due to Weispfenning.
  We follow Kapur et al's notation in \cite{KapurSW13a}.
  
  \begin{definition}[Comprehensive Gr\"obner System]\label{def:cgs}
    Let $I$ be an ideal in $S$ generated by a finite set
    $F\subseteq S$ and $\LL$ be a an algebraically closed field
    containing $\KK$. Assume that $V_1, W_1,\dots,V_r,W_r$ are
    varieties in $\LL^n$, and $G_1$, $\dots$, $G_r$ 
    are finite sets of polynomials in $S$.  A set of tripiles
    $\mathcal{G}=\{ (V_1,W_1,G_1),\dots, (V_r,W_r,G_r)\}$ is called a
    comprehensive Gr\"obner system of $I$ on
    $V = \bigcup_{i=1^r} V_i \setminus W_i$, if for every $a \in V$
    and every specialisation $\sigma_a$ of $S$, $\sigma_a(G_i)$ is a
    Gr\"obner basis of $\sigma_a(I)$ in $\LL[X]$ when $a$ is in
    $V(V_i)\setminus V(W_i)$, for $i=1,\dots,r$.  If $V=\LL^m$, we
    simply call $\mathcal{G}$ a comprehensive Gr\"obner system of
    $I$. Each $(V_i,W_i,G_i)$ is called a branch of $\mathcal{G}$.  A
    comprehensive Gr\"obner system $\mathcal{G}$ of $I$ is called
    faithful, if every element of $G_i$ is in $I$.
  \end{definition}

  \begin{definition}[Comprehensive Gr\"obner Basis]\label{def:cgs}
    Let $I$ be an ideal in $S$ and $\LL$ be an algebraically closed
    field containing $\KK$. Assume that $V$ is a subset of $\LL^m$. A
    finite subset $G$ of $I$is called a comprehensive Gr\"obner basis
    of $I$ on $V$, if for all specialisations
    $\sigma_a :R \rightarrow \LL$ of $S$, where $a \in V$, the set
    $\sigma_a(G)$ is a Gr\"obner basis of the ideal generated by
    $\sigma_a(I)$ in $\LL[X]$.  If $V=\LL^m$, we simply call $G$ a
    comprehensive Gr\"obner basis of $I$. A comprehensive Gr\"obner
    basis $G$ of $I$ is called faithful, if every element of $G$ is in
    $I$.
  \end{definition}

  Having defined comprehensive Gr\"obner bases, Weispfenning proved
  the existence of a comprehensive Gr\"obner basis for every ideal in
  $S$ \cite{Weispfenning92}. In the latter reference, he gave a
  non-constructive proof first, and an algorithm later.

  Following the first algorithm proposed by Weispfenning, algorithms
  for computing a comprehensive Gr\"obner basis essentially construct
  a faithful comprehensive Gr\"obner system
  $\mathcal{G}=\{ (V_1,W_1,G_1),\dots, (V_r,W_r,G_r)\}$. Then the
  union $G=\cup_{i=1}^r G_i$ will be a comprehensive Gr\"obner
  basis. Roughly speaking, the varieties $V_i$ and $W_i$ are typically
  obtained by considering the monomials that are vanished by
  specialisations, and simultaneously, using a Gr\"obner basis
  computation algorithm, a Gr\"obner basis under the conditions
  imposed by the specialisations is computed. Below we present a
  modified version of Kapur, et al.'s algorithm by Dehghani and
  Hashemi from \cite{HashemiDB17}. ``Other cases'' in line 16 of the
  algorithm refers to those cases that the Gr\"obner basis is
  $1$. Dehghani and Hashemi group all those cases together with the
  aim of speeding the computations up.  In line 13, $\MDBasis$
  computes a minimal Dickson basis for a given set of polynomials in
  $S$. For more details, refer to \cite{HashemiDB17}.

\begin{algorithm}
  \caption{$\PGBMAIN$\label{alg:pgbmain}}
  \begin{algorithmic}[1]
    \REQUIRE 1.~~$N, W \subseteq \KK[U]$ finite;\quad
    2.~~$F\subseteq\KK[U][X]$ finite\smallskip
    \ENSURE
    $PGB$ a Gr\"obner system of $F$ on $V(N)\setminus V(W)$ 
    \smallskip

    \STATE{$PGB:=\emptyset$}
    \IF{$V(N)\setminus V(W) = \emptyset$}
    \RETURN{$\emptyset$}
    \ENDIF
    \STATE{$G:=\RGB(F \cup N,<)$}
    \IF{$1 \in G$}
    \RETURN{$\{(N,W,\{1\})\}$}
    \ENDIF
    \STATE{$G_r:=G\cap \KK[U]$}
    \IF{$V(G_r) \setminus V(W) = \emptyset$}
    \RETURN{$PGB$}
    \ELSE
    \STATE{$G_m:=\MDBasis(G\setminus G_r)$\\
    $h=lcm(h_1,\dots,h_k)$ with $h_i=LC_{<_1}(g_i)$ for each $g_i \in
    G_m$
    \IF{$V(G_r) \setminus V(W\times \{h\}) \ne \emptyset$}
    \STATE{$PGB:=PGB \cup \{G_r, W\times \{h\},G_m\}$}
    \ENDIF
    \RETURN{$PGB \cup \bigcup_{h_i \in \{h_1,\dots,h_k\}} \PGBMAIN(G_r
      \cup \{h_i\}, W \times \{h_1h_2\dots h_{i-1}\}, G\setminus G_r)
      \cup \{(\text{Other Cases}, \{1\})\}$}
  }
    \ENDIF
  \end{algorithmic}
\end{algorithm}

 
\section{Testing Binomiality of Chemical Reaction Networks Using
  Comprehensive Gr\"obner Systems}\label{sec:crn-cgs}

  In this section we present computations on biochemical networks,
  using comprehensive Gr\"obner bases, in order to test binomiality of
  the corresponding steady state ideals.

  In
  \cite{eisenbud-sturmfels-binomials,craciun_toric_2009,millan_structure_2018},
  the authors call an ideal binomial if there exists a basis for the
  ideal whose polynomials have at most two terms. In particular, as it
  is discussed in the latter references, one can see that an ideal is
  binomial if and only if its reduced Gr\"obner bases with respect to
  every term order is binomial. Our definition of binomiality is as in
  \cite{DBLP:conf/synasc/RahkooyM20,DBLP:conf/casc/RahkooyRS20}, which
  is slightly different from
  \cite{eisenbud-sturmfels-binomials,craciun_toric_2009,millan_structure_2018}.
  We call an ideal binomial if there exists a basis for the ideal
  whose polynomials have exactly two terms. That is, we do not
  consider monomials in the basis. Similar to the definition of
  binomiality in
  \cite{eisenbud-sturmfels-binomials,craciun_toric_2009,millan_structure_2018},
  one can easily observe that, for the case of our definition, an
  ideal is binomial if and only if its reduced Gr\"obner bases with
  respect to every term order is binomial. In terms of parametric
  polynomial rings, i.e., $\KK[U][X]$, we discuss the binomiality
  using a comprehensive Gr\"obner system. That is in particular the
  case for the steady state ideals of chemical reaction networks.
  
  As computing a comprehensive Gr\"obner basis is doen via computing
  the branches of a comprehensive Gr\"obner system, we basically
  compute the latter and check the binomiality of the Gr\"obner basis
  at each branch. Then a comprehensive Gr\"obner basis of a steady
  state ideal will be binomial if and only if the Gr\"obner basis at
  each branch of a comprehensive system is binomial. One can consider
  the generic comprehensive Gr\"obner bases, introduced in
  \cite{Weispfenning92}, however as it is mentioned in the latter
  reference, computing a generic comprehensive Gr\"obner basis is not
  feasible in practice.

  In this paper, for our computations on the steady state ideals of
  the chemical reaction networks, we consider $\LL=\bar{\KK}$, the
  algebraic closure of $\KK$. In practice, for the computation
  purpose, the coefficient field is considered to be $\QQ$,
  extended by the parameters, i.e., $\QQ(k_1,\dots,k_m)$; hence the
  comprehensive Gr\"obner system computations are carried out over
  $\QQ(k_1,\dots,k_m)[x_1,\dots,x_n]$.
  
  Our computations are carried out via version 4.2.0 of the computer
  algebra system Singular
  \cite{singular}\footnote{\url{http://www.singular.uni-kl.de}}, the
  grobcov package (whose latest version is available at A. Montes'
  website)\footnote{\url{https://mat.upc.edu/en/people/antonio.montes}}.
  For instructions on the grobcov package we refer the reader to the
  book \cite{grobcov-book} and examples by A. Montes.  We have done
  fully automated computations on sets of examples, in particular on
  biochemical models from the BioModels' repository
  \cite{BioModels2015a}~\footnote{\url{https://www.ebi.ac.uk/biomodels}}.
  Our computations have been done on a 2.48 MHz AMD EPYC 7702 64-Core
  Processor in a Debian GNU/Linux 10 machine with 211 gB memory.

\subsection{$n$-Site Phosphorylation}

Multisite phosphorylation–dephosphorylation cycles or $n$-site
phosphorylations (for $n \in \NN$) are studied by Wang and Sontag in
\cite{Wang-Sontag-fautile-cycle2008} in terms of
multi-stationarity. P\'erez-Mil\'an et al. in
\cite{perez_millan_chemical_2012} have shown that for every
$n \in \NN$, $n$-site phosphorylation has a binomial steady state. As
mentioned earlier, in the latter reference, the authors did not take
into account the specialisations of the constant rates.  In this
subsection, we first do some reductions on a basis of the steady state
ideal of $n-$phosphorylations and prove its binomiality. This
essentially gives us the unconditional binomiality of
$n-$phosphorylation, defined and investigated in
\cite{DBLP:conf/casc/RahkooyRS20,DBLP:conf/synasc/RahkooyM20}. Our
algebraic maniplations below are simple and avoid the criterion
presented by P\'erez-Mil\'an et al. in
\cite{perez_millan_chemical_2012}.

Using Wang and Sontag's notation in
\cite{Wang-Sontag-fautile-cycle2008} for the variables and parameters,
for a fixed positive integer $n$, the $n$-site phosphorylation
reaction network is the following.
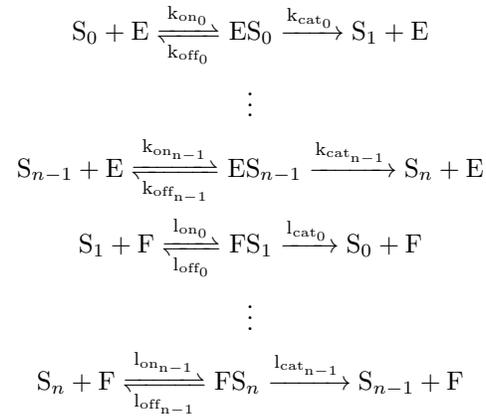
\begin{gather*}
  \ce{S_0 + E <=>[k_{\text{on}_0}][k_{\text{off}_0}] ES_0
    ->[k_{\text{cat}_0}] S_1 + E} \\
  \vdots\\
  \ce{S_{n-1} + E <=>[k_{\text{on}_{n-1}}][k_{\text{off}_{n-1}}] ES_{n-1}
    ->[k_{\text{cat}_{n-1}}] S_n + E} \\
  \ce{S_1 + F <=>[l_{\text{on}_0}][l_{\text{off}_0}] FS_1
    ->[l_{\text{cat}_0}] S_0 + F} \\
  \vdots\\
  \ce{S_n + F <=>[l_{\text{on}_{n-1}}][l_{\text{off}_{n-1}}] FS_n
    ->[l_{\text{cat}_{n-1}}] S_{n-1} + F} 
\end{gather*}

The parameters of the reaction network are
$k_{\text{on}_0},\dots, k_{\text{on}_{n-1}}$,
$k_{\text{off}_0},\dots,k_{\text{off}_{n-1}}$,
$k_{\text{cat}_0},\dots, k_{\text{cat}_{n-1}}$,
$l_{\text{on}_0},\dots, l_{\text{on}_{n-1}}$,
$l_{\text{off}_0},\dots,l_{\text{off}_{n-1}}$,
$l_{\text{cat}_0},\dots, l_{\text{cat}_{n-1}}$.  Let the variables
$s_0,\dots,s_n, c_0,\dots, c_{n-1}, d_1,\dots, d_n, e, f$ represent
the concentrations of the species
$\ce{S_0},\dots, \ce{S_n}, \ce{ES_0},\dots, \ce{ES_{n-1}},
\ce{FS_1},\dots, \ce{FS_n}, \ce{E}, \ce{F}$ respectively.  The ODEs
describing change of the concentrations of the species for this
reaction network are the following:
 \begin{align*}
   \dot{s_0} = P_0 = & -k_{\text{on}_0} s_0e + k_{\text{off}_0} c_0 +
                       l_{\text{cat}_0} d_1,\\ 
   \dot{s_i} = P_i = & -k_{\text{on}_i} s_i e + k_{\text{off}_i} c_i +
                       k_{\text{cat}_{i-1}} 
                       c_{i-1} -l_{\text{on}_{i-1}} s_i f +
                       l_{\text{off}_{i-1}} d_i + l_{\text{cat}_i} d_{i+1},  \\
                     & i=1,\dots,n-1,\\
   \dot{c_j} = Q_j = & k_{\text{on}_j} s_j e -
                       (k_{\text{off}_j}+k_{\text{cat}_j}) c_j, \quad 
                       j=0,...,n-1,\\
   \dot{d_k} = R_k= & l_{\text{on}_{k-1}} s_k
                      f-(l_{\text{off}_{k-1}}+l_{\text{cat}_{k-1}}
                      ) d_k, \quad k=1,...,n.
 \end{align*}

 The ODEs for $s_n, e$ and $f$ are linear combinations of the above
 ODEs, hence they are redundant and we skip them in this article.

 In order to show unconditional binomiality of the steady state ideal
 of $n$-phosphorylation, we perform reductions on the the generators
 of the steady state ideal so that a binomial basis is obtained. First
 of all, note that polynomials $Q_j$ and $R_k$ are already
 binomial. Reducing $P_0$ with respect to $Q_0$, we obtain
 \begin{align*}
   P_0'= & P_0+Q_0 \\
   = & -k_{\text{on}_0} s_0e + k_{\text{off}_0} c_0 +
       l_{\text{cat}_0} d_1 \\
         & + k_{\text{on}_0} s_0 e -
           (k_{\text{off}_0}+k_{\text{cat}_0}) c_0\\
   = & l_{\text{cat}_0} d_1 + k_{\text{cat}_0} c_0,
 \end{align*}
 which is a binomial.

 Now we reduce $P_i$ with respect to $P_0'$, $Q_j$ and $R_k$ as
 follows. First we reduce $P_i$ with respect to $R_I$
 \begin{align*}
   P_i+R_i =& \\
            & -k_{\text{on}_i} s_i e + k_{\text{off}_i} c_i + k_{\text{cat}_{i-1}}
              c_{i-1} -l_{\text{on}_{i-1}} s_i f + l_{\text{off}_{i-1}} d_i + l_{\text{cat}_i}
              d_{i+1}\\
            &+ l_{\text{on}_{i-1}} s_i
              f-(l_{\text{off}_{i-1}}+l_{\text{cat}_{i-1}}
              ) d_i\\
   =& -k_{\text{on}_i} s_i e + k_{\text{off}_i} c_i + k_{\text{cat}_{i-1}}
      c_{i-1} + l_{\text{cat}_i}d_{i+1} -l_{\text{cat}_{i-1}} d_i.
 \end{align*}
 Then we reduce the result with respect to $Q_i$
 \begin{align*}
   P_i+R_i+Q_i =& \\
                & -k_{\text{on}_i} s_i e + k_{\text{off}_i} c_i + k_{\text{cat}_{i-1}}
                  c_{i-1} + l_{\text{cat}_i}d_{i+1}
                  -l_{\text{cat}_{i-1}} d_i\\
                & + k_{\text{on}_i} s_i e -  (k_{\text{off}_i}+k_{\text{cat}_i}) c_i  \\
   =& k_{\text{cat}_{i-1}}
      c_{i-1} + l_{\text{cat}_i}d_{i+1}
      -l_{\text{cat}_{i-1}} d_i +k_{\text{cat}_i} c_i. 
 \end{align*}
 For $i=1$, the above can be reduced with respect to $P_0'$
 \begin{align*}
   P_1' = P_1+R_1+Q_1-P_0'= &
                              k_{\text{cat}_{0}}  c_{0} + l_{\text{cat}_1}d_{1}
                              -l_{\text{cat}_{0}} d_1 +k_{\text{cat}_1} c_1  \\
                            & -\left( l_{\text{cat}_0} d_1 + k_{\text{cat}_0} c_0 \right)\\
   =& l_{\text{cat}_1}d_{1} + k_{\text{cat}_1} c_1,
 \end{align*}
 which is a binomial.

 Similarly, for $i=2,\dots,n$, $P_i$ can be reduced to a binomial with
 respect to $R_i$, $Q_I$ and $P_{i-1}'$. Therefore, a binomial basis
 can be obtained this way for the steady state ideal.

 As the algebraic manipulations above do not consider into account the
 specialisations of the parameters, we computed comprehensive
 Gr\"obner system of the steady state ideals for the cases $n=1,2$ to
 test the binomiality under specialisations. $1$-site phosphorylation
 and $2$-site phosphorylations have been studied in
 \cite{perez_millan_chemical_2012} using the criteria presented in
 that article as well.

 \begin{Example}[1-site phosphorylation,
   \cite{perez_millan_chemical_2012}, Example 2.1]
   \begin{gather*}
     \ce{S_0 + E <=>[k_{\text{on}_0}][k_{\text{off}_0}] ES_0
       ->[k_{\text{cat}_0}] S_1 + E} \\
     \ce{S_1 + F <=>[l_{\text{on}_0}][l_{\text{off}_0}] FS_1
       ->[l_{\text{cat}_0}] S_0 + F}.
   \end{gather*}

   Let the variables representing the change of the concentrations of
   the species $\ce{S_0}$, $\ce{S_1}$, $\ce{ES_0}$, $\ce{FS_1}$,
   $\ce{E}$, $\ce{F}$ be $s_0$, $s_1$, $c_0$, $d_1$, $e$, $f$
   respectively, and let the parameters be
   $k_{\text{on}_0}, k_{\text{off}_0}, k_{\text{cat}_0},
   l_{\text{on}_0}, l_{\text{off}_0}, l_{\text{cat}_0}$.
   
   The steady state ideal for 1-site phosphorylation reaction is
   generated by
   \begin{align*}
     \dot{s_0} = & -k_{\text{on}_0} s_0 e + k_{\text{off}_0} c_0 +
                   l_{\text{cat}_0} d_1, \\ 
     \dot{s_1}=  & -k_{\text{on}_1} s_1 e + k_{\text{off}_1} c_1 +
                   k_{\text{cat}_0} c_0 - l_{\text{on}_0}s_1f +
                   l_{\text{off}_0}d_1,\\ 
     \dot{c_0}=&  k_{\text{on}_0} s_0 e -( k_{\text{off}_0} +
                 k_{\text{cat}_0}) c_0,\\ 
     \dot{d_1}=&  l_{\text{on}_0} s_1 f -( l_{\text{off}_0} +
                 l_{\text{cat}_0}) d_1. 
   \end{align*}
   We skip the ODEs for $e$ and $f$ as they are linear combination of
   the other ODEs. Renaming the variables as
   \[ s_0 = x_1,s_1 = x_2,c_0 = x_3,d_1 = x_4,e = x_5,f = x_6, \] we
   computed the comprehensive Gr\"obner system for the steady state
   ideal using Singular. It contains 25 branches, out of which 6 are
   binomial. We recall that in this article, a binomial ideal is an
   ideal that is generated by a set of binomials (not including
   monomials).  For the last branch, $V_{25}$ and $W_{25}$ are the
   zero sets of the following sets of polynomials in
   $\QQ[k_{\text{on}_0}, k_{\text{off}_0}, k_{\text{cat}_0},
   l_{\text{on}_0}, l_{\text{off}_0},
   l_{\text{cat}_0}][x_1,\dots,x_6]$ respectively:
   \begin{align*}
     & \{ l_{\text{cat}_0}, k_{\text{on}_0}\},\\
     & \{ k_{\text{off}_0} k_{\text{cat}_0}
       l_{\text{on}_0}+k_{\text{cat}_0}^2l_{\text{on}_0} \}. 
    \end{align*}
    The corresponding Gr\"obner basis is
    \begin{align*}
      \{f_1=& k_{\text{cat}_0} x_3, \\
      f_2=&l_{\text{on}_0} x_2x_6-l_{\text{off}_0} x_4 \},
    \end{align*}
    which obviously is not binomial.
    
    An example of a branch with binomial Gr\"obner basis is branch 24,
    for which $V_{24}$ and $W_{24}$ are the zero sets of the following
    sets, respectively:
       \begin{align*}
     & \{ k_{\text{off}_0}+k_{\text{cat}_0}, k_{\text{on}_0}\},\\
     & \{ l_{\text{on}_0} k_{\text{cat}_0} \}. 
    \end{align*}
    The corresponding Gr\"obner basis is
    \begin{align*}
      \{f_1=& k_{\text{cat}_0} x_3 +l_{\text{cat}_0} x_4, \\
      f_2=&l_{\text{on}_0} x_2x_6+(-l_{\text{off}_0}-l_{\text{cat}_0}) x_4 \}.
    \end{align*}
 \end{Example}

 \begin{Example}[2-site phosphorylation,
   \cite{perez_millan_chemical_2012}, Example 3.13]
   The steady state ideal for the 2-site phosphorylation reaction is
   generated by
   \begin{align*}
     \dot{s_0} = P_0 = & -k_{\text{on}_0} s_0e + k_{\text{off}_0}
                            c_0 + l_{\text{cat}_0} d_1,\\
     \dot{s_1} = P_1 = & -k_{\text{on}_1} s_1 e + k_{\text{off}_1} c_1
                         + k_{\text{cat}_{0}} 
                         c_{0} -l_{\text{on}_{0}} s_1 f +
                         l_{\text{off}_{0}} d_1 + l_{\text{cat}_1}  d_{2},  \\
     \dot{c_0} = Q_0 = & k_{\text{on}_0} s_0 e -
                         (k_{\text{off}_0}+k_{\text{cat}_0}) c_0,\\ 
     \dot{c_0} = Q_1 = & k_{\text{on}_1} s_1 e -
                         (k_{\text{off}_1}+k_{\text{cat}_1}) c_1,\\ 
     \dot{d_1} = R_1= & l_{\text{on}_{0}}
                        s_1f-(l_{\text{off}_{0}}+l_{\text{cat}_{0}} 
                        ) d_1, \\
     \dot{d_2} = R_2= & l_{\text{on}_{1}}
                        s_2f-(l_{\text{off}_{1}}+l_{\text{cat}_{1}}  ) d_2 ,
   \end{align*}
   where the variables are
   \[ s_0, s_1, s_2, c_0, c_1, d_1, d_2, e, f \]
   and the parameters   are
   \[
     k_{\text{on}_0},k_{\text{on}_1},k_{\text{off}_0},k_{\text{off}_1},k_{\text{cat}_0},k_{\text{cat}_1},l_{\text{on}_0},l_{\text{on}_1},l_{\text{off}_0},l_{\text{off}_1},l_{\text{cat}_0},l_{\text{cat}_1}.\]
   We have computed a comprehensive Gr\"obner system for this system
   using Singular. It has 1187 branches, out of which 36 are
   binomial. The last branch of the comprehensive Gr\"obner system is
   as follows. $V_{1187}$ is the zero set of
   $l_{\text{off}_1}+l_{\text{cat}_1}$ and $W_{1187}$ is the zero set
   of the following polynomial:
   \begin{align*}
     &
       k_{\text{on}_0}k_{\text{on}_1}k_{\text{off}_0}k_{\text{off}_1}
       k_{\text{cat}_0}k_{\text{cat}_1}l_{\text{on}_0}l_{\text{on}_1}l_{\text{off}_0}l_{\text{cat}_0}l_{\text{cat}_1} \\ 
      & +k_{\text{on}_0}k_{\text{on}_1}k_{\text{off}_0}k_{\text{off}_1}k_{\text{cat}_0}k_{\text{cat}_1}l_{\text{on}_0}l_{\text{on}_1}l_{\text{cat}_0}^2l_{\text{cat}_1} \\
      &+k_{\text{on}_0}k_{\text{on}_1}k_{\text{off}_0}k_{\text{cat}_0}k_{\text{cat}_1}^2l_{\text{on}_0}l_{\text{on}_1}l_{\text{off}_0}l_{\text{cat}_0}l_{\text{cat}_1} \\
      &+k_{\text{on}_0}k_{\text{on}_1}k_{\text{off}_0}k_{\text{cat}_0}k_{\text{cat}_1}^2l_{\text{on}_0}l_{\text{on}_1}l_{\text{cat}_0}^2l_{\text{cat}_1} \\
       &+k_{\text{on}_0}k_{\text{on}_1}k_{\text{off}_1}k_{\text{cat}_0}^2k_{\text{cat}_1}l_{\text{on}_0}l_{\text{on}_1}l_{\text{off}_0}l_{\text{cat}_0}l_{\text{cat}_1}\\
      &+k_{\text{on}_0}k_{\text{on}_1}k_{\text{off}_1}k_{\text{cat}_0}^2k_{\text{cat}_1}l_{\text{on}_0}l_{\text{on}_1}l_{\text{cat}_0}^2l_{\text{cat}_1} \\
       &+k_{\text{on}_0}k_{\text{on}_1}k_{\text{cat}_0}^2k_{\text{cat}_1}^2l_{\text{on}_0}l_{\text{on}_1}l_{\text{off}_0}l_{\text{cat}_0}l_{\text{cat}_1}\\
      &+k_{\text{on}_0}k_{\text{on}_1}k_{\text{cat}_0}^2k_{\text{cat}_1}^2l_{\text{on}_0}l_{\text{on}_1}l_{\text{cat}_0}^2l_{\text{cat}_1}.
   \end{align*}
   Renaming the variables  as
   \[ s_0 = x_1,s_1 = x_2,s_2 = x_3,c_0 = x_4,c_1 = x_5,d_1 = x_6,d_2
     = x_7,e = x_8,f = x_9,\] the Gr\"obner basis for every
   specialisation of the parameters in $V_{1187}\setminus W_{1187}$ is
   the following.
   \begin{align*}
     f_1=&k_{\text{cat}_1}x_5-l_{\text{cat}_1}x_7, \\
     f_2= &k_{\text{cat}_0}x_4-l_{\text{cat}_0}x_6,\\
      f_3= &l_{\text{on}_1}x_3x_9,\\
     f_4= &l_{\text{on}_0}x_2x_9+(-l_{\text{off}_0}-l_{\text{cat}_0})x_6,\\
     f_5= &(k_{\text{on}_1}l_{\text{off}_0}+k_{\text{on}_1}l_{\text{cat}_0})x_6x_8+(-k_{\text{off}_1}l_{\text{on}_0})x_5x_9+(-l_{\text{on}_0}l_{\text{cat}_1})x_7x_9,\\
     f_6=&(k_{\text{on}_1})x_2x_8+(-k_{\text{off}_1})x_5+(-l_{\text{cat}_1})x_7,\\
     f_7= &(k_{\text{on}_0})x_1x_8+(-k_{\text{off}_0})x_4+(-l_{\text{cat}_0})x_6,\\
     f_8= &(l_{\text{on}_1}l_{\text{off}_0}+l_{\text{on}_1}l_{\text{cat}_0})x_3x_6,\\
      f_9=
         &(k_{\text{on}_1}k_{\text{off}_0}k_{\text{cat}_1}l_{\text{cat}_0}
     +k_{\text{on}_1}k_{\text{cat}_0}k_{\text{cat}_1}l_{\text{cat}_0})x_2x_6\\
      &+(-k_{\text{on}_0}k_{\text{off}_1}k_{\text{cat}_0}l_{\text{cat}_1}-k_{\text{on}_0}k_{\text{cat}_0}k_{\text{cat}_1}l_{\text{cat}_1})x_1x_7,\\
     f_{10}=&(k_{\text{on}_0}k_{\text{off}_1}l_{\text{on}_0}l_{\text{cat}_1}+k_{\text{on}_0}k_{\text{cat}_1}l_{\text{on}_0}l_{\text{cat}_1})x_1x_7x_9\\
     &+(-k_{\text{on}_1}k_{\text{off}_0}k_{\text{cat}_1}l_{\text{off}_0}-k_{\text{on}_1}k_{\text{off}_0}k_{\text{cat}_1}l_{\text{cat}_0})x_4x_6 \\
     & +(-k_{\text{on}_1}k_{\text{cat}_1}l_{\text{off}_0}
       l_{\text{cat}_0}-k_{\text{on}_1}k_{\text{cat}_1}l_{\text{cat}_0}^2)x_6^2,\\
     f_{11}=&(
       k_{\text{on}_0}k_{\text{off}_1}k_{\text{cat}_0}l_{\text{on}_1}l_{\text{off}_0}l_{\text{cat}_1}+k_{\text{on}_0}k_{\text{off}_1}k_{\text{cat}_0}l_{\text{on}_1}l_{\text{cat}_0}l_{\text{cat}_1} \\
     &  +k_{\text{on}_0}k_{\text{cat}_0}k_{\text{cat}_1}l_{\text{on}_1}l_{\text{off}_0}l_{\text{cat}_1}
     +k_{\text{on}_0}k_{\text{cat}_0}k_{\text{cat}_1}l_{\text{on}_1}l_{\text{cat}_0}l_{\text{cat}_1})
       x_1x_3x_7.
     \end{align*}
     One can observe that the above branch of the comprehensive
     Gr\"obner system is not binomial.
 \end{Example}

 We carried on the computations for comprehensive Gr\"obner system of
 the steady state ideal of $n-$phosphorylation for $n=2, 3, 4, 5$ in
 Singular with the time limit of six hours. The results of the
 computations are summarised in Table \ref{n-phosph-table}.  In this
 table, DNF refers to did not finish.

 {
   \setlength\LTleft{0pt} \setlength\LTright{0pt}
  \begin{longtable}{@{\extracolsep{\fill}}|c|c|c|c|c|}
    \caption{Comprehensive Gr\"obner System of
    n-Phosphorylations}\label{n-phosph-table}\\ 
    \hline
    &  $\#$branches & $\#$binomial branches & $\%$ of binomial
                                              branches & time(s)\\
    \hline
    \endhead  
    $2-$phosph.& 1187& 36 & 3.03 &24 \\
    $3-$phosph. &57857 & 216 & 0.37 & 2231\\
    $4-$phosph. & - & - & - & DNF\\
    $5-$phosph. & - & -& -& DNF\\
    \hline
  \end{longtable}
}

As the number of variables and parameters grow drastically when $n$
increases, comprehensive Gr\"obner system computations did not finish
in a reasonable time period for $n\geq4$.

We also computed a comprehensige Gro\"obner system of
2-phosphorylation in Maple, using Dehghani and Hashemi's PWWG
package\footnote{\url{https://amirhashemi.iut.ac.ir/sites/amirhashemi.iut.ac.ir/files//file_basepage/pggw_0.txt}},
which uses a modification of Kapur et al.'s algorithm so that the
branches with Gr\"obner basis $\{1\}$ are ignored
\cite{HashemiDB17}. According to the authors' experiments in
\cite{HashemiDB17}, this modification results in speed-up of the
computations. However, even for 2-phosphorylation the computations did
not finish in six hours in Maple.

As we see from the computations in this subsection, there are several
branches of teh $n-$phosphorylations that are not binomial. This means
that for certain values of the rate constants, $n-$phosphorylation is
not binomial, while the computations without taking into account the
specialisations of the rate constants leads to the binomiality.

\subsection{BioModels}
Our main benchmark for computing comprehensive Gr\"obner system of
steady state ideals, are the biochemical models from the BioModels
repository \cite{BioModels2015a}, which is typically used for such
computations. As a first example, we present biomodel 629 and the
corresponding computations in the following example.
 
 \begin{Example}[BIOMD0000000629, \cite{BioModels2015a}]
  The corresponding ODEs for biomodel 629 are the following;
  \begin{align*}
    \dot{x_1}=& -k_2x_1x_3 + k_3x_2, \\
    \dot{x_2}=&k_2x_1x_3 - k_3x_2 - k_4x_2x_4 + k_5x_5, \\
    \dot{x_3}=& -k_2x_1x_3 + k_3x_2, \\
    \dot{x_4}=& -k_4x_2x_4 + k_5x_5,\\
    \dot{x_5}=&k_4x_2x_4 - k_5x_5,
  \end{align*}
where $k_1,\dots,k_5$ are the parameters and $x_1,\dots,x_5$ are the
variables. Comprehensive G\"obner system computation over the ring
$\QQ[k_1,\dots,k_5][x_1,\dots,x_5]$ in Singular results in 10 branches
with the following conditions and Gr\"obner bases.

{
  \setlength\LTleft{0pt}
  \setlength\LTright{0pt}
  \begin{longtable}{@{\extracolsep{\fill}}|c|c|c|c|}
    \caption{Comprehensive Gr\"obner System of
    BIOMD0000000629}\label{bio629}\\ 
    \hline
     branch  &  $V$ & $W$ & GB\\
    \hline
    \endhead
    1 & $0$ & $k_2k_4$ & $k_4x_2x_4-k_5x_5, k_2x_1x_3-k_3x_2$ \\
    2 &$k_4$ & $k_2k_5$ & $k_5x_5, k_2x_1x_3-k_3x_2$ \\
    3 & $k_5,k_2$ & $k_2$ & $k_2x_1x_3-k3x_2$ \\
    4 & $k_5,k_4,k_2$ & $k_3$ & $k_3x_2$ \\
    5 & $k_5,k_4,k_3,k_2$ & $1$ & $0$ \\
    6 & $k_4,k_2$ & $k_3k_5$ & $k_5x_5,k_3x_2$ \\
    7 & $k_4,k_3,k_2$ & $k_5$ & $k_5x_5$ \\
    8 & $k_2$ & $k_5,k_4,k_3$ & $k_3k_5x_5, k_3x_2$ \\
    9 & $k_3, k_2$ & $k_4$ & $k_4x_2x_4-k_5x_5$ \\
    10 & $k_5,k_2$ & $k_3k_4$ & $k_3x_2$ \\    
    \hline
  \end{longtable}
}

There are three branches with binomial Gr\"obner basis for biomodel
629. All the branches have either monomial or binomial Gr\"obner
basis.
\end{Example}

In Table \ref{biocgs-table}, we present the results of our
computations for some biomodels from the Biomodels repository
\cite{BioModels2015a}. As computing comprehensive Gr\"obner system of
systems with large number of variables is very expensive, we have
considered those biomodels that have relatively small number of
species (correspondingly, relatively small number of variables), so
that the computations took less than ten minutes for those biomodels.
In Table \ref{biocgs-table}, one can find the number of branches of
the corresponding comprehensive Gr\"obner systems, the number of
branches that are binomial, and their percentage. Except for biomodels
271 and 519 that have no binomial branch, all other biomodels have at
least one binomial branch. For two biomodels (283 and 486), at least
half of their branches are binomial.

The largest biomodel we have considered is the model 26. We note that
this model is a MAPK reaction network. It has been studied in
\cite{EnglandErrami:17b}, where the authors associated a graph to the
CRN and used a trick based on vertex cover in order to reduce the
number of the polynomials in the steady state ideal into 2
polynomials.

{
  \setlength\LTleft{0pt}
  \setlength\LTright{0pt}
  \begin{longtable}{@{\extracolsep{\fill}}|c|c|c|c|}
    \caption{Branches of Comprehensive Gr\"obner Systems of
    Biomodels}\label{biocgs-table}\\ 
    \hline
    model  & $\#$branches & $\#$binomial branches & $\%$ of binomial branches \\
    \hline
    \endhead  
    26 & 46870 & 164 & 0.35 \\
    40 & 35 & 6 & 17.00 \\
    92 & 10 & 4 & 40.00 \\
    101 & 81 & 11 & 13.40 \\
    104 & 4 &  1 & 25.00 \\
    156 & 25 & 5 & 20.00 \\
    159 & 36 & 6 & 16.66 \\
    178 & 24& 2 & 8.33 \\
    194 & 19 & 5 & 26.31 \\
    233 & 18 & 5 & 27.78 \\
    267 & 12& 2 &16.67 \\
    271& 92& 0 & 0.00 \\
    272 & 44 & 7& 15.91 \\
    282& 18& 4 & 22.22 \\
    283 & 2 & 1 & 50.00 \\
    289 & 351 & 43 & 12.25 \\
    321& 26& 5& 19.23 \\
    363& 15& 2& 13.33 \\
    459& 40& 9& 22.50 \\
    486& 3& 2& 66.67 \\
    519& 128& 0& 0.00 \\
    546& 15& 1& 6.67 \\
    629& 10& 4 & 40.00\\
    \hline
  \end{longtable}
}

\section{Conclusion}\label{sec:conclusion}

We address the problem of binomiality of the steady state ideal of a
chemical reaction network. The binomiality problem has been widely
considered in the literature of mathematics and chemical reaction
network theory and is still an active research area.  Finding
binomiality and toricity is a hard problem from both a theoretical and
a practical point of view. The computational methods typically rely on
Gr\"obner bases..

The authors have recently investigated binomiality and toricity in
several papers. We have given efficient algorithms for testing
toricity in \cite{grigoriev2019efficiently}. We also have considered
the binomiality from a first-order logic point of view and gave
efficient computational results and studied biomodels systematically
via quantifier elimination
\cite{sturm2020firstorder,grigoriev2019efficiently}. Other than those,
we have considered the concept of unconditional binomiality, which
considers rate constants as variables, and gave polynomial time linear
algebra and graph theoretical approaches for detecting binomiality
\cite{DBLP:conf/synasc/RahkooyM20,DBLP:conf/casc/RahkooyRS20}.

The existing work on binomiality of steady state ideals do not take
into account the effect of assigning values to the rate constats
during the computations. In the present work, we consider the problem
of binomiality when the parameters can be specialised. Our approach to
this parametric binomiality problem is naturally based on
comprehensive Gr\"obner bases.  We make systematic computations on
$n-$phosphorylations and biomodels and detect the branches of the
Gr\"obner systems that are binomial.
Our computations via comprehensive Gr\"obner systems show that in
several cases, the comprehensive Gr\"obner bases for steady state
ideals are not binomial, while using other methods, e.g., considering
rate constants as variables or doing computations without considering
the effect of specialisation, one may consider those steady state
ideal as binomial ideals.

As in this paper the concept of comprehensive Gr\"obner bases is used
for the first time on chemical reaction network theory, we propose
using this approach for studying further properties of chemical
reaction networks.

\subsection*{Acknowledgments}
This work has been supported by the interdisciplinary bilateral
project ANR-17-CE40-0036/DFG-391322026 SYMBIONT
\cite{BoulierFages:18a,BoulierFages:18b}. We would like to thank
A. Hashemi and M. Dehghani for the discussions on comprehensive
Gr\"obner bases and providing us with their Maple package.

\bibliography{references}
\bibliographystyle{splncs04}
\end{document}